# Hyperspectral imaging of excitons within a moiré unit-cell with a sub-nanometer electron probe

Sandhya Susarla[1,2#], Mit H. Naik[1,3#], Daria D. Blach[4], Jonas Zipfel[2], Takashi Taniguchi[6], Kenji Watanabe[7], Libai Huang[4], Ramamoorthy Ramesh[1,3,8], Felipe H. da Jornada[5,9], Steven G. Louie[1,3*], Peter Ercius[2*], Archana Raja[2*]

1: Materials Science Division, Lawrence Berkeley National Laboratory, Berkeley, CA, 94720, USA
2: Molecular Foundry, Lawrence Berkeley National Laboratory, Berkeley, CA, 94720, USA
3: Department of Physics, University of California, Berkeley, CA, 94720, USA
4: Department of Chemistry, Purdue University, West Lafayette, IN, 47907, USA
5: Department of Materials Science and Engineering, Stanford University, Stanford, CA, 94305, USA
6: International Center for Materials Nanoarchitectonics, National Institute for Materials Science, 1-1 Namiki, Tsukuba, 305-0044, Japan
7: Research Center for Functional Materials, National Institute for Materials Science, 1-1 Namiki, Tsukuba, 305-0044, Japan
8: Department of Materials Science and Engineering, University of California, Berkeley, CA, 94720, USA
9: Stanford PULSE Institute, SLAC National Accelerator Laboratory, Menlo Park, CA, 94025, USA

# Equal Contribution
*Corresponding authors
  Email: araja@lbl.gov, percius@lbl.gov, sglouie@berkeley.edu

**Abstract:**

Electronic and optical excitations in two-dimensional moiré systems are uniquely sensitive to local atomic registries, leading to materials- and twist-angle specific correlated electronic ground states with varied degree of localization. However, there has been no direct experimental correlation between the sub-nanometer structure and emergent excitonic transitions, comprising tightly-bound pairs of photoexcited electrons and holes. Here, we use cryogenic transmission electron microscopy and spectroscopy to simultaneously image the structural reconstruction and associated localization of the lowest-energy intralayer exciton in a rotationally aligned heterostructure of $WS_2$ and $WSe_2$ monolayers. In conjunction with optical spectroscopy and *ab initio* calculations, we determine that the exciton center-of-mass wavefunction is strongly modulated in space, confined to a radius of ~ 2 nm around the highest-energy stacking site in the moiré unit-cell, forming a triangular lattice. Our results provide direct evidence that atomic reconstructions lead to the strongly confining moiré potentials and that engineering strain at the nanoscale will enable new types of excitonic lattices.

**Introduction:**

Moiré superlattices formed by stacking monolayers of van der Waals crystals are a burgeoning platform for discovery of fundamental physical phenomena (*1*). For example, the moiré superlattice of semiconducting transition metal dichalcogenides (TMDCs) have been predicted to form a topologically protected lattice of bound electron-hole pairs or excitons that can act as a model system for quantum simulations and technologies (*2–4*). Recent optical spectroscopy studies have found signatures of interlayer and intralayer moiré excitons in TMDC heterostructures (*5–8*), investigated exciton diffusion in a superlattice potential (*9*), and found evidence for the cooperative nature of moiré excitons (*10*). Nanoscale modulation or confinement of the center of mass of *interlayer* excitons has been recently probed in momentum space using ultrafast angle-resolved photoemission electron spectroscopy (ARPES) (*11*). However, the real-space center-of-mass confinement within a moiré unit cell and the long-range ordering of the excitonic lattice in real materials has not been visualized so far due to the diffraction limit of optical probes or due to the loss of phase information in ARPES. Such real-space visualization of the exciton localization is necessary to address the fundamental question of whether a moiré superlattice can support a periodic array of well-localized quantum excitations (*2–4*).

Electron microscopy can be used to measure structure and electronic transitions at high resolution. Using TMDC heterostructures as model systems, previous studies have measured *either* their atomic scale structural reconstruction by annular dark field-scanning transmission electron microscopy (ADF- STEM) (*12*, *13*) *or* the excitonic signatures using spatially averaged STEM- low-loss electron energy loss spectroscopy (EELS) (*14–16*). The advent of high sensitivity direct electron detectors incorporated into EEL spectrometers along with cryogenic holders and monochromation provides an unprecedented opportunity to simultaneously probe the weak exciton signals and the structural reconstruction at the sub-nanometer scale.

In this work, we use simultaneous ADF-STEM and low-loss STEM-EELS mapping with hyperspectral analysis to directly image the in-plane structural reconstruction of a rotationally aligned (R-stacked) $WS_2/WSe_2$ moiré superlattice and the corresponding amplitude of the moiré exciton center-of-mass wavefunction within the moiré unit cell. Our ADF-STEM measurements and theoretical calculations reveal that the lattice-mismatched moiré superlattice undergoes a large in-plane structural reconstruction for a system governed by relatively weak interlayer van der Waals interactions. We find that the area of the low-energy chalcogen-metal stacking configurations ($B^{Se/W}$ and $B^{W/S}$) increases while that of the higher energy chalcogen-chalcogen stacking (AA) decreases compared to the unrelaxed structure. Furthermore, we show that the center of mass wavefunction of the lowest energy intralayer moiré exciton is localized within a ~2 nm radius of the AA stacking site and forms a triangular lattice, in good agreement with our first-principles *GW* plus Bethe-Salpeter equation (*GW*-BSE) calculations. The ability to spatially probe excitons at high resolution can be extended to other systems and is the first step to design periodic localized excitations.

**Results:**

We prepared a rotationally aligned (R-stacked) heterostructure of $WS_2$ and $WSe_2$ monolayers encapsulated within ~25nm thick hexagonal boron nitride (hBN) on each side. The sample was suspended over a 20x20 $\mu m^2$ hole in a custom fabricated silicon TEM grid. hBN creates a uniform dielectric environment that narrows excitonic linewidths (*17*) and protects the sample from oxidation and beam damage. The twist angle between the layers is verified using position averaged convergent beam electron diffraction (PACBED) and second harmonic generation (SHG) (Figure S1-S2), and the emergence of moiré excitons is confirmed by cryogenic optical reflectance contrast and photoluminescence measurements (Figure S3). We then collected hyperspectral images from multiple regions under cryogenic conditions (100 K) using simultaneous ADF-STEM-EELS with a ~100 meV monochromated electron beam focused to less than a nanometer (Figure 1A, Supplementary Information, Figure S4, S5). We implemented a custom data analysis routine based on unit cell averaging to improve the signal-to-noise ratio in the structural image and final EEL spectra and map (Supplementary Information, Figure S6-16).

$WS_2$ and $WSe_2$ monolayers are direct band gap semiconductors with a lattice parameter of 3.18 Å and 3.32 Å, respectively. The 4.4% lattice mismatch gives rise to a moiré periodicity of ~8 nm for near-zero relative twist. Figure 1B shows a schematic of the heterostructure with three high symmetry stacking configurations: AA has metal and chalcogens vertically stacked, $B^{Se/W}$ has Se stacked on W, and $B^{W/S}$ has W stacked on S. The latter two are Bernal stacking regions. Due to steric effects, our DFT calculations show that stacking in the moiré superlattice affects the local energy landscape, and AA stacking has the highest energy (Figure 1C) (*18–20*).

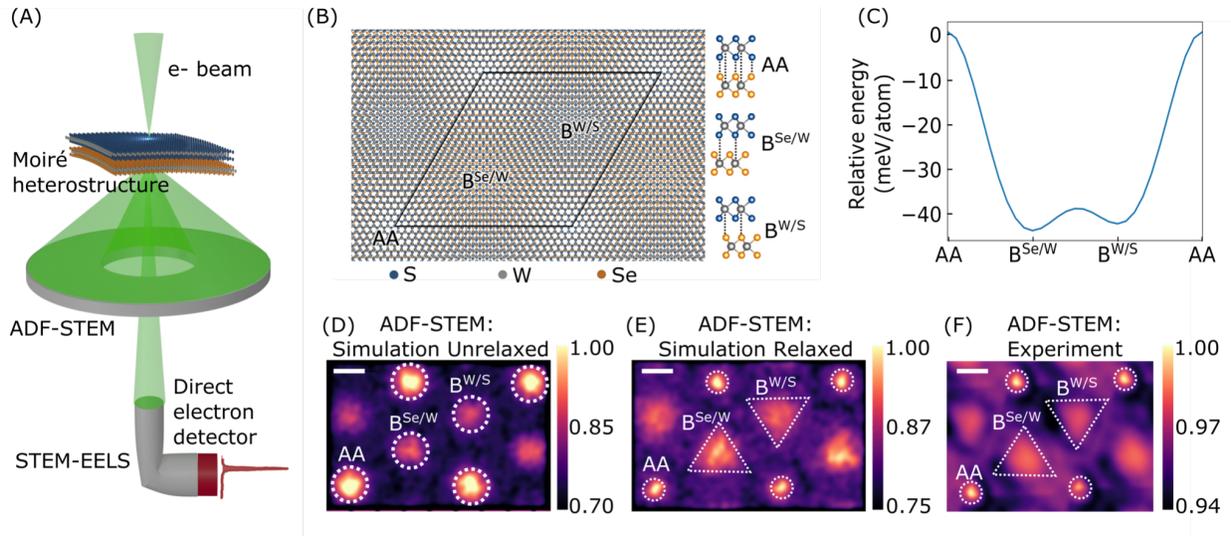

**Figure 1: Structural reconstruction of a WS$_2$/WSe$_2$ moiré heterostructure.** (A) Schematic describing the simultaneous monochromated ADF-STEM-EELS experiment with a direct electron detector. (B) Structural model of reconstructed R-stacked WS$_2$/WSe$_2$ heterostructure showing different stacking configurations in a moiré unit cell. (C) Relative energy of the various stacking configurations in the moiré superlattice. (D, E) ADF-STEM multislice simulations of unrelaxed and relaxed heterostructures. (F) Experimental ADF-STEM unit cell periodically repeated to match the relaxed heterostructure in D and E. Scale bars (D-F): 2 nm. The color bar (D-F) indicates the normalized intensity.

We obtain sub-nano-scale structural information by analyzing ADF-STEM images, where intensity is proportional to the locally summed atomic number (Z) at each atomic column along the electron beam direction (Materials and Methods, Supplementary Information). Quantitative ADF-STEM has been used in the past to study in-plane reconstructions in large-period, lattice-matched moiré systems such as twisted bilayer MoS$_2$, WS$_2$, and MoS$_2$/WS$_2$ (*12*). To help interpret our experiments, we simulated ADF-STEM images of the theoretically predicted unrelaxed and relaxed moiré superlattice using multislice simulations (see Supplementary Information) (*21*). Figure 1D displays the simulation of the unrelaxed moiré heterostructure where circular regions of higher intensity correspond to the AA, B$^{Se/W}$, and B$^{W/S}$ stacking arrangements. The AA region ($Z_{WW} = 142$) appears brighter than the B$^{Se/W}$ ($Z_{Se/W} = 58$) and B$^{W/S}$ ($Z_{W/S} = 54$) regions. Note that, without accounting for a structural reconstruction, all the stacking configurations have roughly the same area (Figure S17). On the other hand, the ADF-STEM simulation of the relaxed moiré heterostructure (Figure 1E) shows clear differences among the sizes of the bright and dark regions. Driven by the stacking energy landscape (Figure 1C), the effective area corresponding to the high-energy AA configuration is reduced as compared to the unrelaxed structure, and the $B^{Se/W}$ and $B^{W/S}$ regions form larger, triangular domains. Experimentally, we imaged an area of the sample containing many moiré unit-cells with a 5 Å probe step size. While direct atomic resolution imaging was challenging due to the overlap of atoms with very small projected distances (Figure S18), our simulations allowed us to differentiate AA sites from B sites. We averaged the ADF-STEM intensities over 162 moiré unit cells (see Materials and Methods, Figure S6) to obtain the final image shown in Figure 1F. The experimental ADF-STEM unit cell is consistent across scans (Figure S7), not affected by hBN (Figure S19), and in agreement with the multislice simulation of the relaxed heterostructure (Figure 1E), providing the first direct evidence of in-plane structural reconstructions in a $WS_2/WSe_2$ moiré superlattice.

The structural transformation we observe (Figure 1C) leads to a strain redistribution over the whole moiré unit-cell (with a maximum strain of ~1%) in the individual layers due to a fine balance between the strain energy cost and stacking energy gain (*22*) (Figure S20). The experimental in-plane structural reconstruction we present here are distinct from those in previous first-principle calculations (*20*, *23*) and ADF-STEM reports (*12*) in twisted lattice-matched systems, where the strain was localized to the boundary between the lowest-energy stacking configurations. Usually, when a lattice-mismatched TMDC heterostructure is formed by thermodynamics-based growth techniques such as chemical vapor deposition, the heterostructure does not reconstruct (*9*). The surprisingly large in-plane structural reconstruction we observe in the mechanically stacked van der Waals interfaces was recently predicted by theory, and the associated inhomogeneous strain in the individual layers is responsible for the formation of flat electronic bands (*22*) and spatially modulated excitonic states (*24*).

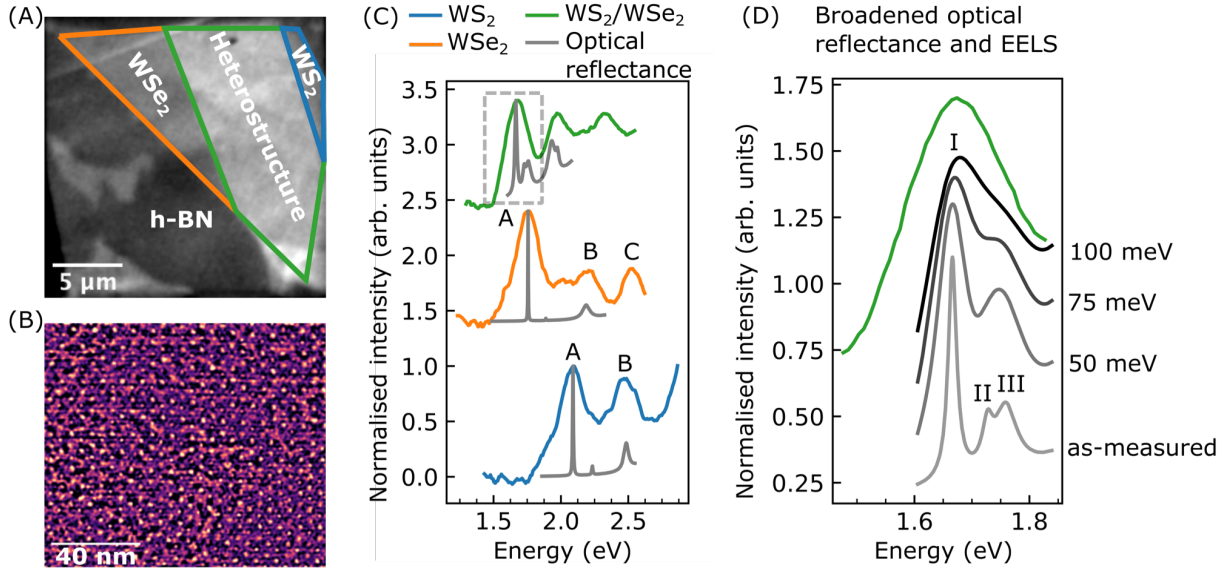

**Figure 2: Comparison of EELS with optical absorption.** (A) Optical image of the R-stacked $WS_2/WSe_2$ heterostructure transferred on the TEM grid with hBN encapsulation. (B) False colored band-pass filtered ADF-STEM image acquired during STEM-EELS showing a uniform moiré pattern over a 100 x 100 $nm^2$ region. (C) Spatially averaged EEL spectra of the moiré heterostructure (green) compared to monolayer $WSe_2$ (orange) and $WS_2$ (blue) with reference optical reflectance spectra (gray). (D) Zoomed spectra of the moiré heterostructure showing the three distinct moiré exciton peaks resolvable by optics measurements. Gaussian broadening of 100 meV matches the EELS resolution.

We study the influence of the in-plane structural reconstruction on the moiré excitonic states through ensemble optical spectroscopy and localized low-loss STEM-EELS on the same sample, where we get higher spectral resolution in the former and higher spatial resolution in the latter. Figure 2A shows the optical image of the R-stacked $WS_2/WSe_2$ heterostructure embedded within hBN. Figure 2B shows a representative band-pass filtered ADF-STEM image of the heterostructure displaying a moiré pattern over a 100x100 $nm^2$ region with 0.5 nm resolution. Figure 2C shows the average EEL spectra acquired from regions of monolayer $WSe_2$ (orange), monolayer WS$_2$ (blue) and WS$_2$/WSe$_2$ (green) as compared to optical reflectivity spectra (gray). EELS background subtraction was performed using Gauss-Lorentz fitting in the pre- and post-edge onset of the exciton peak (*25*) and smoothed by Savitzsky-Golay algorithm (Figure S8-S12). Similar to optical absorption, the EEL signal is proportional to the imaginary part of the dielectric constant of the material (*26*). The split A and B exciton peak positions arising from spin-orbit coupling of both WS$_2$ (A: 2.10 eV; B: 2.40 eV) and WSe$_2$ (A: 1.75 eV; B: 2.10 eV) monolayers in low loss EELS match well with optical data, suggesting that the two spectroscopic techniques display a similar oscillator strength for the main exciton peaks. Figure 2D shows the zoomed-in moiré exciton peaks. The lowest energy peak (1.65 eV) is slightly red shifted compared to the WSe$_2$ monolayer A exciton peak (1.75 eV). The optical reflectivity measurement shows the three lowest-energy intralayer moiré exciton peaks (labeled I, II and III). While we can only resolve this fine structure of the moiré exciton peaks optically, subsequent broadening of the optical spectra by 100 meV yields good agreement with the EEL spectra (Figure 2D, Table S1). The most significant spectral contribution (58%) in the 1.6-1.7 eV range comes from the moiré exciton peak I (Figure S21), consistent with the largest oscillator strength predicted by *ab initio GW*-BSE calculations. Interlayer excitons are not observed in optical reflectance or EELS due to their negligible (~1000x smaller) oscillator strengths compared to the intralayer excitations (*27*).

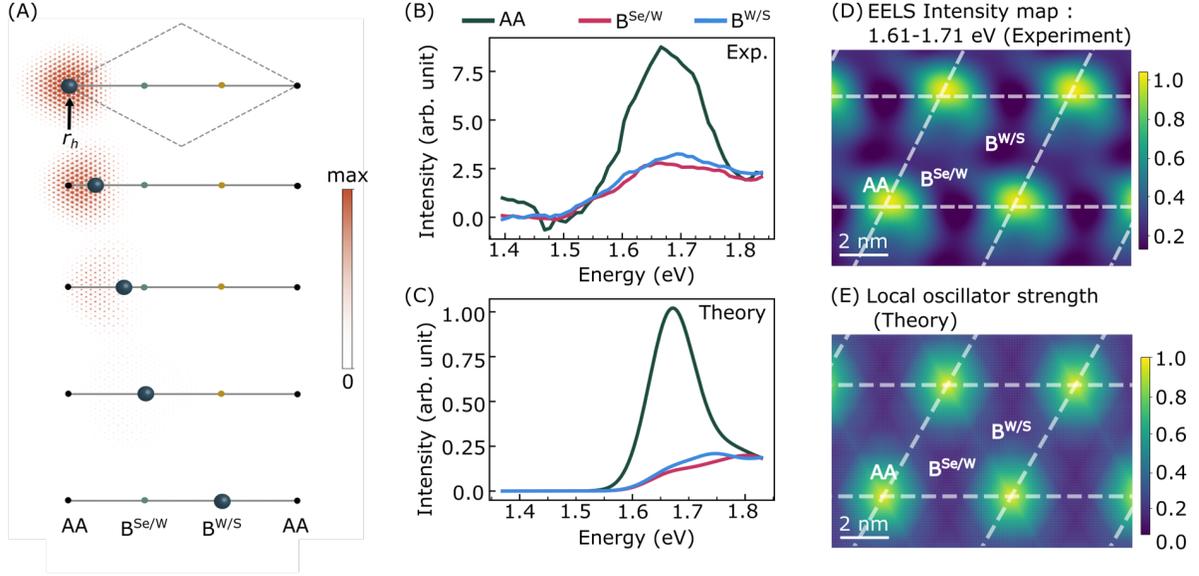

**Figure 3: Sub-Nanoscale spatial mapping of moiré excitons:** (A) Calculated electron charge density contributing to the peak I exciton, $\psi_S(r_e, r_h)$, for a fixed position of the hole, $r_h$, in the moiré superlattice. The evolution of the electron charge density is plotted by moving the hole position across the longer diagonal of the superlattice, showing the spatially modulated character of the peak I exciton. (B and C) Comparison of spatially resolved experimental EELS and $GW$-BSE theory for different stacking configurations AA, $B^{Se/W}$ and $B^{W/S}$. (D) Experimental unit cell averaged and tiled EELS map obtained by a rolling average with a 1.3 nm radius at different unit cell locations showing the modulation of the EELS intensity which is primarily composed of peak I at 1.65 eV (E) Theoretically calculated oscillator strength distribution of the peak I exciton with similar radial blurring as the experiment.

The splitting of the WSe$_2$ A exciton peak into three peaks in the WS$_2$/WSe$_2$ moiré superlattice is predicted (*24*) to be a result of a strong in-plane structural reconstruction observed in ADF-STEM measurements (Figure 1), unlike other moiré systems where layer hybridization can also contribute to the moiré-exciton fine structure (*7*). Our first-principles $GW$-BSE calculations reproduce the three peaks observed in the optical reflection contrast spectrum (Figure S22). The wave function describing a moiré exciton state *S* is a linear combination of valence-to-conduction transitions in the Brillouin zone: $\psi^S(\boldsymbol{r_e}, \boldsymbol{r_h}) = \sum_{vck} A^S_{vck} \phi_{ck}(\boldsymbol{r_e}) \phi^*_{vk}(\boldsymbol{r_h})$,, where $\boldsymbol{r_e}$ ($\boldsymbol{r_h}$) is the electron (hole) coordinate, $A^S_{vck}$ is the exciton envelope function, $\phi_{nk}$ are the single-particle Bloch states with band *n* and wavevector *k*, *v* and *c* label valence and conduction bands, respectively, and *S* is an exciton band index. Furthermore, our calculations show that the exciton associated with peak I forms a spatially modulated Wannier exciton with maximum charge density at the AA stacking in the moiré superlattice (Figure 3A). We compare the oscillator strength of the moiré exciton, primarily peak I, with the experimental data by applying a 1.3 nm hard mask around the AA, $B^{Se/W}$, and $B^{W/S}$ positions determined from ADF-STEM structural images (Supplementary information; Figure S13, S15, S16). While the electron beam is spatially confined to less than a nanometer in diameter, the low-loss EELS delocalization limits the achievable spatial resolution to about 1 nm (*26*, *28*). We observe a clear difference in the peak intensities derived from the three stacking regions, indicating that the excitonic peak we observe in EELS (primarily peak I) has an inhomogeneous oscillator strength in the moiré unit-cell with the maximum intensity at AA stacking regions (Figure 3B and Figure S15,16). This is in very good agreement with the results from our first-principles *GW*-BSE calculations constraining the exciton-photon coupling matrix element $\langle S | \int d^3 r \Psi^\dagger(r) A(r) \cdot \mathbf{v} \, \Psi(r) | 0 \rangle$ to a specific spatially resolved region by taking the vector potential $A(r)$ to have a Gaussian profile, and where **v** is the velocity operator, $\Psi(r)$ is a field operator, and $|0\rangle$ is the ground state wavefunction (Figure 3C, supplementary information). The oscillator strength of peak I has the largest contribution from around the AA stacking since the photoexcited electron and hole charge density are maximum in this region (Figure 3A). For a Wannier-type exciton, the oscillator strength distribution is proportional to $\psi^S(r_e = x, r_h = x)$ which corresponds to the wavefunction of the exciton center-of-mass coordinate, $\boldsymbol{R} = \frac{r_e + r_h}{2} = \boldsymbol{x}$, and relative coordinate $\boldsymbol{r} = \frac{r_e - r_h}{2} = 0$, i.e. $\psi^S(\boldsymbol{R} = \boldsymbol{x}, \boldsymbol{r} = 0)$.

To determine the spatial extent of the exciton modulation within a moiré unit cell, we generated a hyperspectral unit cell by shifting a 1.3 nm mask to different unit cell locations across a 100x100 nm² scan, effectively creating a rolling average (Supplementary Information,

Figure S13, S14). Figure 3D shows the unit-cell averaged and periodically repeated EELS intensity map generated by summing spectral intensity from 1.61 - 1.71 eV (see Figure S16, 23-24 for more details). We observe that the maximum intensity of the exciton is constrained to the AA site, which forms a triangular lattice in the moiré pattern. Furthermore, the exciton intensity rapidly reduces beyond ~2nm from the AA site, in good agreement with the corresponding *GW*-BSE calculated oscillator strength map of the moiré exciton peak I (Figure 3E).

The effect of structural reconstruction in moiré lattices has been predicted to generate strong electronic and excitonic confinement (*22*, *24*). By taking advantage of the spatial and spectral resolution in cryogenic correlated ADF-STEM and STEM-EELS techniques, we directly capture the sub-nanometer scale in-plane structural reconstruction and the corresponding exciton localization without the limitations of diffractive optics. This reconstruction is not observed in thermodynamically grown heterostructures (*9*). We find that the structural reconstruction is robust over a sub-micron area of the sample, in spite of potential sources of disorder introduced by the multilayer mechanical stacking process. The in-plane structural reconstruction leads to spatially modulated electronic bands (*22*, *29*) in this system, corroborated by the recent observations of strongly correlated Wigner crystal and Mott insulating phases (*30*, *31*) .

The exciton wavefunction as a function of the center-of-mass coordinate (with small electron-hole relative coordinate) is localized to the AA site within a radius of ~2 nm, forming a triangular lattice. The diminished exciton intensity at the Bernal stacking regions (Figure 3D, E) is in agreement with the surprising absence of trion formation upon hole doping of the same heterostructure system (*5*, *24*). Unit-cell averaging, which samples over hundreds of square nanometers, greatly improves the signal to noise ratio of the EELS intensity map. This strongly suggests that the moiré potential is preserved over a large area of the sample and supports the formation of a triangular lattice of excitons.

We discover that the strong atomic relaxations within each moiré cell lead to confinement of excitons at a particular stacking site. This opens up the possibility of nanoscopic engineering of bosonic lattices in moiré heterostructures through external strain, twist-angle and number of layers.

**Acknowledgments:** The authors thank Dr. Medha Dandu and Prof. Tony Heinz for thoughtful discussions on the results and Johan Carlström for his inputs on sample fabrication. The authors also thank Dr. Jim Ciston and Dr. Chenyu Song for their support during the STEM-EELS experiments. We acknowledge Prof. Crozier for useful discussions about background subtraction in low-loss EELS.

**Funding:** Work at the Molecular Foundry was supported by the Office of Science, Office of Basic Energy Sciences, of the U.S. Department of Energy under Contract No. DE-AC02-05CH11231. S.S. and R.R. are supported by the DOE Quantum Materials Program. A. R. acknowledges support through the Early Career LDRD Program of Lawrence Berkeley National Laboratory under DOE Contract No. DE-AC02-05CH11231. DDB acknowledges support from the U.S. Department of Energy, Office of Science, Office of Workforce Development for Teachers and Scientists, Office of Science Graduate Student program through the SCGSR program administered by the Oak Ridge Institute for Science and Education (ORISE) for the DOE. ORISE is managed by ORAU under contract number DE-SC0014664. J. Z. acknowledges funding by the Deutsche Forschungsgemeinschaft (DFG, German Research Foundation) through the Walter-Benjamin Program 462503440, and through the U.S. Department of Energy, Office of Science, Basic Energy Sciences, Materials Sciences and Engineering Division, award number DE-SC0022289. The theoretical calculations were primarily supported by the Director, Office of Science, Office of Basic Energy Sciences, Materials Sciences and Engineering Division of the US Department of Energy under the van der Waals heterostructure program (KCWF16), contract number DE-AC02-05CH11231 which provided the PUMP approach for the exciton properties of moiré systems. The work was also supported by the Center for Computational Study of Excited-State Phenomena in Energy Materials (C2SEPEM) at LBNL, funded by the U.S. Department of Energy, Office of Science, Basic Energy Sciences, Materials Sciences and Engineering Division under Contract No. DE-AC02-05CH11231, as part of the Computational Materials Sciences Program which provided advanced codes and simulations, and supported by the Theory of Materials Program (KC2301) funded by the U.S. Department of Energy, Office of Science, Basic Energy Sciences, Materials Sciences and Engineering Division under Contract No. DE-AC02-05CH11231 which provided conceptual analyses of the moiré excitons. Computational resources were provided by National Energy Research Scientific Computing Center (NERSC), which is supported by the Office of Science of the US Department of Energy under contract no. DE-AC02-05CH11231, Stampede2 at the Texas Advanced Computing Center (TACC), The University of Texas at Austin through Extreme Science and Engineering Discovery Environment (XSEDE), which is supported by National Science Foundation under grant no. ACI-1053575 and Frontera at TACC, which is supported by the National Science Foundation under grant no. OAC-1818253. K.W. and T.T. acknowledge support from JSPS KAKENHI (Grant Numbers 19H05790, 20H00354 and 21H05233). L.H. acknowledges support from US Department of Energy, Office of Basic Energy Sciences, through award DE-SC0016356.

**Author contributions:**

    Project supervision: AR, PE, RR, SGL

Conceptualization: AR, PE, SS, MHN

ADF-STEM-EELS experiments and data analysis: SS, PE

*GW*-BSE theory: MHN, SGL, FHJ

Sample fabrication: DDB, TT, KW

Optical Reflectivity, Second Harmonic Generation and data analysis: JZ, DDB

Scientific discussion: AR, RR, SGL, FHJ, LH, PE, SS, MHN

Writing – original draft: SS, MHN, AR, PE, FHJ

Writing – review & editing: SS, MHN, AR, PE, FHJ, SGL, RR, LH, DDB, JZ

**Competing interests:** Authors declare that they have no competing interests.

**Data and materials availability:** All data are available in the main text or the supplementary materials.

**Supplementary Materials**

Materials and Methods

Supplementary Text

Figs. S1 to S24